\documentclass[10pt]{article}
\usepackage{amsmath}
\usepackage{amsfonts}
\usepackage{amssymb}
\usepackage{amsthm}
\usepackage{graphicx}

\numberwithin{equation}{section}
\renewcommand{\vec}{\boldsymbol}
\renewcommand{\leq}{\leqslant}
\renewcommand{\geq}{\geqslant}
\newcommand{\bigO}{\mathcal{O}}
\newcommand{\rme}{\mathrm{e}}
\newcommand{\rmd}{\mathrm{d}}
\newcommand{\eref}[1]{(\ref{#1})}
\newcommand{\case}[2]{{\textstyle \frac{#1}{#2}}}
\newcommand{\etal}{{\it et~al\/}\ }

\theoremstyle{definition}
\newtheorem{theorem}{Theorem}[section]
\newtheorem{corollary}{Corollary}[section]
\newtheorem{proposition}{Proposition}[section]
\newtheorem{conjecture}{Conjecture}[section]

\begin{document}

\noindent {\Large\bf Asymptotic Regimes of Magnetic Bianchi
                     Cosmologies}\\[2ex plus 0.3ex minus 0.3ex]
\noindent {\bf Joshua T.\ Horwood\footnotemark[1] and
               John Wainwright\footnotemark[1]}
\footnotetext[1]{Department of Applied Mathematics, University of Waterloo,
Waterloo, Ontario, Canada N2L~3G1}
\addtocounter{footnote}{1}

\begin{quote} \small
  \textbf{Abstract.} We consider the asymptotic dynamics of the
  Einstein-Maxwell field equations for the class of non-tilted Bianchi
  cosmologies with a barotropic perfect fluid and a pure homogeneous
  source-free magnetic field, with emphasis on models of Bianchi type
  VII$_{0}$, which have not been previously studied. Using the orthonormal
  frame formalism and Hubble-normalized variables, we show that, as is the
  case for the previously studied class~A magnetic Bianchi models, the magnetic
  Bianchi VII$_{0}$ cosmologies also exhibit an oscillatory approach to the
  initial singularity. However, in contrast to the other magnetic Bianchi
  models, we rigorously establish that typical magnetic Bianchi VII$_{0}$
  cosmologies exhibit the phenomena of asymptotic self-similarity breaking and
  Weyl curvature dominance in the late-time
  regime.\\[1.5ex plus 0.3ex minus 0.3ex]
  \noindent \textbf{Key words.} Non-tilted magnetic Bianchi cosmologies.
\end{quote}


\section{Introduction} \label{sec_intro}

The influence of an intergalactic magnetic field on cosmological models has
been investigated for over four decades both from a theoretical and
observational point of view. Cosmologists speculate that such a field could
be primordial in origin, that is, one that came into existence at the Planck
time. Observational techniques rely on studying processes such as the
temperature distribution of the cosmic microwave background radiation (CMBR),
primeval nucleosynthesis and the Faraday rotation of linearly polarized
radiation emitted from extragalactic radio sources. Barrow \etal \cite{B97}
derive an upper bound of \mbox{$B_{0} < 3.4 \times 10^{-9}\,
(\Omega_{0}h_{50}^{2})^{1/2}\, \mathrm{gauss}$} on the present strength of any
spatially homogeneous primordial magnetic field\footnote{For comparison,
the strength of the Earth's magnetic field at the surface is approximately
$0.5\,\mathrm{gauss}$.} based on data from the COBE satellite ($\Omega_{0}$
is the present value of the density parameter and $h_{50}$ is the Hubble
constant in units of $50\,\mathrm{km\,s^{-1}\,Mpc^{-1}}$). All observations to
date only place an upper bound on the strength of such a magnetic field and
hence are inconclusive as regards its existence.

Any cosmological model which contains a magnetic field is necessarily
anisotropic, since isotropy is violated by the preferred direction of the
magnetic field vector. Consequently, one must analyze the Einstein field
equations in models more general than the homogeneous and isotropic
Friedmann-Lema\^{\i}tre (FL) models. The simplest family of cosmological models
that can admit a magnetic field are the so-called Bianchi cosmologies, that is,
models which admit a three-parameter group of isometries acting orthogonally
transitively on spacelike hypersurfaces. The models are thus spatially
homogeneous, but, in general, anisotropic. We assume that the models contain
a barotropic perfect fluid whose four-velocity is orthogonal to the group
orbits, and that observers comoving with the fluid measure a pure source-free
magnetic field. We also assume that the perfect fluid satisfies an equation
of state $p = (\gamma - 1)\mu$, where $\gamma$ is constant and satisfies
$\frac{2}{3} < \gamma < 2$, the cases $\gamma = 1$ (dust) and $\gamma =
\frac{4}{3}$ (radiation) being of primary interest. We shall refer to solutions
of the combined Einstein-Maxwell field euqations that satisfy the above
properties as \textit{magnetic Bianchi cosmologies}. It is known that the field
equations lead to restrictions on the Bianchi-Behr type of the isometry group,
namely, that it is of types I, II, VI$_{0}$ or VII$_{0}$ (in class~A) or
type III (in class~B)\footnote{We refer to Ellis and MacCallum \cite{EM69} for
this terminology.}.

Significant progress has been made in the study of magnetic Bianchi
cosmologies. Collins \cite{C72} was the first to use techniques from dynamical
systems theory to obtain qualitative results concerning the evolution of
axisymmetric Bianchi I models under the assumption that the magnetic field is
aligned along a shear eigenvector. More recently, LeBlanc \etal \cite{LKW95}
gave a qualitative analysis of the dynamics of magnetic Bianchi cosmologies of
type VI$_{0}$ in their asymptotic regimes, that is, near the initial
singularity and at late times. This work made use of the orthonormal frame
formalism of Ellis and MacCallum \cite{EM69} and Hubble-normalized variables
(see \cite{WE}, ch.~5 and 6). This formalism was also used in \cite{L97} and
\cite{L98} to give a similar analysis of magnetic Bianchi universes of types I
and II. Most recently, Crowe \cite{C98} extended the results to magnetic
Bianchi models of type III. There remains one class which has not been
previously analyzed, namely magnetic Bianchi cosmologies of type VII$_{0}$.

Our goal in this paper is to fill this gap by giving a qualitative analysis of
the dynamics of magnetic Bianchi cosmologies of type VII$_{0}$ in their
asymptotic regimes. Bianchi cosmologies of type VII$_{0}$ are of interest
because they represent anisotropic generalizations of the flat FL models. The
asymptotic dynamics of the non-magnetic models at late times has only been
analyzed in detail relatively recently (see~\cite{WHU99}). It is worth
comparing non-magnetic Bianchi cosmologies of group type VII$_{0}$ with their
counterparts of group type I, which are also anisotropic generalizations of
the flat FL models. The non-magnetic Bianchi type I cosmologies are
asymptotically self-similar at late times, that is, they are approximated by
a self-similar solution at late times. This self-similar solution is in fact
the flat FL solution, which means that the Bianchi I cosmologies undergo
asymptotic isotropization. In contrast, for values of the equation of state
parameter $\gamma$ satisfying $\frac{2}{3} < \gamma < 2$, the Bianchi
VII$_{0}$ cosmologies are not asymptotically self-similar at late times.
Nevertheless, for values of $\gamma$ satisying $1 \leq \gamma \leq
\frac{4}{3}$, they undergo a subtle form of isotropization: the rate of
expansion isotropizes, but the intrinsic gravitational field, as described by
the Weyl curvature tensor, does not. This phenomenon has been referred to as
\textit{Weyl curvature dominance} (see~\cite{WHU99}). One of our specific
goals in this paper is to determine what effect a cosmic magnetic field has
on the above-mentioned isotropization. The method that we use is a
generalization of the analysis of the non-magnetic Bianchi VII$_{0}$ and VIII
models given in \cite{WHU99} and \cite{HHTW03}, respectively.

The plan of paper is as follows. In section~\ref{sec_eveqn}, we present the
evolution equations for the magnetic Bianchi cosmologies of type VII$_{0}$
using the orthonormal-frame formalism and Hubble-normalized variables.
Section~\ref{sec_limits} contains the main result concerning the dynamics in
the late-time regime, namely theorem~\ref{theorem_limits} and
corollary~\ref{cor_limits_scalars}, which give the limits of the
Hubble-normalized variables and of certain physical dimensionless scalars,
thereby describing the asymptotic dynamics at late times. In
section~\ref{sec_past}, we examine the singular asymptotic regime and show
that typical models exhibit an oscillatory singularity. We conclude in
section~\ref{sec_discussion} with a discussion of the cosmological
implications of our results and give an overview of the asymptotic dynamics
of the magnetic Bianchi cosmologies, noting that the present paper completes
the picture.

There are three appendices. Appendix~A contains the proof of the fact that
magnetic Bianchi VII$_{0}$ universes are not asymptotically self-similar at
late times. Appendix~B fills in some of the technical details of the proof of
theorem~\ref{theorem_limits}. Finally, in appendix~C we give expressions for
a dimensionless scalar formed from the Weyl curvature tensor in terms of the
Hubble-normalized variables.


\section{Evolution equations} \label{sec_eveqn}

In this section we give the evolution equations for magnetic
Bianchi cosmologies of type VII$_{0}$. As described in \cite{LKW95} (pg.~517),
we use Hubble-normalized variables
\begin{equation}
  (\Sigma_{+},\Sigma_{-},N_{2},N_{3},\mathcal{H}), \label{eq_eveqn_spA}
\end{equation}
defined relative to a group-invariant orthonormal frame $\{ \vec{e}_{a} \}$,
with $\vec{e}_{0} = \vec{u}$, the fluid 4-velocity, which is normal to the
group orbits.

The variables $\Sigma_{\pm}$ describe the shear of the fluid congruence,
the $N_{2,3}$ are spatial connection variables which describe the intrinsic
curvature of the group orbits and the variable $\mathcal{H}$ describes the
magnetic degree of freedom. The magnetic Bianchi VII$_{0}$ cosmologies are
described by the inequalities $\mathcal{H} > 0$ and $N_{2} N_{3} > 0$. Without
loss of generality, we assume
\begin{equation}
  N_{2} > 0, \qquad N_{3} > 0. \label{eq_eveqn_restA}
\end{equation}
It is convenient to define
\begin{equation}
  N_{+} = \case{1}{2}(N_{2} + N_{3}), \qquad
  N_{-} = \case{1}{2\sqrt{3}}(N_{2} - N_{3}), \label{eq_eveqn_Npm}
\end{equation}
and replace \eref{eq_eveqn_spA} by the state vector
\begin{equation}
  (\Sigma_{+},\Sigma_{-},N_{+},N_{-},\mathcal{H}) . \label{eq_eveqn_spB}
\end{equation}
The restrictions \eref{eq_eveqn_restA} become
\begin{equation}
  N_{+} > 0, \qquad N_{+}^{2} - 3 N_{-}^{2} > 0, \qquad
  \mathcal{H} > 0 . \label{eq_eveqn_restB}
\end{equation}

The state variables \eref{eq_eveqn_spA} and \eref{eq_eveqn_spB} are
dimensionless, having been normalized with the Hubble scalar\footnote{On
account of \eref{eq_eveqn_H}, $H$ is related to the rate of volume expansion
$\theta$ of the fundamental congruence according to $H = \frac{1}{3} \theta$.
We note that all variables of LeBlanc \etal \cite{LKW95} are normalized
with $\theta$.} $H$, which is related to the overall length scale $\ell$ by
\begin{equation}
  H = \frac{ \dot{\ell} }{ \ell }, \label{eq_eveqn_H}
\end{equation}
where the overdot denotes differentiation with respect to clock time along
the fundamental congruence. The state variables depend on a dimensionless
time variable $\tau$ that is related to the length scale $\ell$ by
\begin{equation}
  \ell = \ell_{0}\, \rme^{\tau} , \label{eq_eveqn_ell}
\end{equation}
where $\ell_{0}$ is a constant. The dimensionless time $\tau$ is related
to the clock time $t$ by
\begin{equation}
  \frac{\rmd t}{\rmd \tau} = \frac{1}{H}, \label{eq_eveqn_t_tau}
\end{equation}
as follows from equations \eref{eq_eveqn_H} and \eref{eq_eveqn_ell}. In
formulating the evolution equations we require the deceleration parameter
$q$, defined by
\begin{equation}
  q = -\frac{\ell \ddot{\ell}}{\dot{\ell}^{2}} , \label{eq_eveqn_q_defn}
\end{equation}
and the density parameter $\Omega$, defined by
\begin{equation}
  \Omega = \frac{\mu}{3 H^{2}}. \label{eq_eveqn_Omega_defn}
\end{equation}
We also find it convenient to introduce the magnetic density parameter
$\Omega_{h}$, defined analogously by
\begin{equation}
  \Omega_{h} = \frac{\mu_{h}}{3 H^{2}} , \label{eq_eveqn_Omegah_defn}
\end{equation}
where $\mu_{h}$ is the energy density of the magnetic field. We note that
$\mu_{h}$ is given by
\begin{equation}
  \mu_{h} = \case{1}{2}(h_{1}^{2} + h_{2}^{2} + h_{3}^{2}),
    \label{eq_eveqn_muh_defn}
\end{equation}
where the $h_{\alpha}$, $\alpha=1,2,3$, are the components of the magnetic
field intensity relative to the spatial orthonormal frame
$\{\vec{e}_{\alpha}\}$, which has been chosen so that
\begin{equation}
  h_{\alpha} = (h_{1},0,0) \label{eq_eveqn_halpha}
\end{equation}
(see~\cite{LKW95}).

A complete derivation of the evolution equations for the variables
\eref{eq_eveqn_spB}, which arise from the combined Einstein-Maxwell field
equations, is provided in \cite{LKW95} (see section~2). These evolution
equations read\footnote{These evolution equations are essentially the same
as those given in \cite{LKW95} for magnetic Bianchi VI$_{0}$ models, apart
from a numerical factor multiplying $\mathcal{H}^{2}$. The difference between
Bianchi VII$_{0}$ and Bianchi VI$_{0}$ models lies in the restrictions that
define the state space: the quantity $N_{+}^{2} - 3 N_{-}^{2}$ is negative
for Bianchi VI$_{0}$ models, in contrast to \eref{eq_eveqn_restB}.}
\begin{equation}\begin{split}
  \Sigma'_{+} &= (q-2) \Sigma_{+} - 2 N_{-}^{2} + \case{1}{3}
    \mathcal{H}^{2}, \\
  \Sigma'_{-} &= (q-2) \Sigma_{-} - 2 N_{+} N_{-}, \\
  N'_{+} &= (q + 2 \Sigma_{+}) N_{+} + 6 \Sigma_{-} N_{-}, \\
  N'_{-} &= (q + 2 \Sigma_{+}) N_{-} + 2 \Sigma_{-} N_{+}, \\
  \mathcal{H} ' &= (q - 2 \Sigma_{+} - 1) \mathcal{H} ,
\end{split}\label{eq_eveqn_DE_nonpolar}\end{equation}
where
\begin{align}
  q &= 2 ( \Sigma_{+}^{2} + \Sigma_{-}^{2} ) + \case{1}{6} \mathcal{H}^{2} + 
    \case{1}{2}(3\gamma-2) \Omega , \label{eq_eveqn_q_nonpolar} \\
  \Omega &= 1 - \Sigma_{+}^{2} - \Sigma_{-}^{2} - N_{-}^{2} - \case{1}{6}
    \mathcal{H}^{2} , \label{eq_eveqn_Omega_nonpolar}
\end{align}
and ${}'$ denotes differentiation with respect to $\tau$. For future reference
we also note the evolution equation for $\Omega$:
\begin{equation}
  \Omega ' = [ 2q - (3\gamma-2) ] \Omega , \label{eq_eveqn_Omega_DE_nonpolar}
\end{equation}
and the expression for the magnetic density parameter
\begin{equation}
  \Omega_{h} = \case{1}{6} \mathcal{H}^{2}, \label{eq_eveqn_Omegah_nonpolar}
\end{equation}
in terms of the Hubble-normalized magnetic field intensity $\mathcal{H} =
h_{1}/H$, which follows from \eref{eq_eveqn_Omegah_defn},
\eref{eq_eveqn_muh_defn} and \eref{eq_eveqn_halpha}.

The physical requirement $\Omega \geq 0$ in conjunction with
\eref{eq_eveqn_Npm} implies that the variables $\Sigma_{\pm}$, $N_{-}$ and
$\mathcal{H}$ are bounded, but places no restriction on $N_{+}$ itself.
In fact, it will be shown in appendix~A (see proposition~\ref{prop_Nplus})
that if $\Omega > 0$ and $\frac{2}{3} < \gamma < 2$, then for any initial
conditions
\begin{equation}
  \lim_{\tau \rightarrow +\infty} N_{+} = +\infty.\label{eq_eveqn_N_plus_limit}
\end{equation}
The first step in analyzing the dynamics at late times ($\tau \rightarrow
+\infty$) is to introduce new variables which are bounded at late times and
which enable us to isolate the oscillatory behaviour associated with
$\Sigma_{-}$ and $N_{-}$. Motivated by \cite{WHU99}, we define
\begin{equation}
  \Sigma_{-} = R \cos \psi, \qquad
  N_{-} = R \sin \psi, \qquad
  M = \frac{1}{N_{+}}, \label{eq_eveqn_cov}
\end{equation}
where $R \geq 0$.

In terms of the new variables $(\Sigma_{+},R,\mathcal{H},M,\psi)$, the
evolution equations~\eref{eq_eveqn_DE_nonpolar} have the following form
\begin{equation}\begin{split}
  \Sigma'_{+} &= (Q-2)\Sigma_{+} - R^{2} + \case{1}{3} \mathcal{H}^{2} + 
    (1+\Sigma_{+}) R^{2} \cos 2\psi, \\
  R' &= \left[ Q+\Sigma_{+}-1 + (R^{2} - 1 -\Sigma_{+}) \cos 2 \psi 
    \right] R, \\
  \mathcal{H}' &= \left[ Q - 2 \Sigma_{+} - 1 + R^{2} \cos 2\psi 
    \right] \mathcal{H}, \\
  M' &= - \left[ Q + 2 \Sigma_{+} + R^{2} ( \cos 2\psi + 3 M \sin 2 \psi )
    \right] M, \\
  \psi ' &= \frac{1}{M} \left[2 + (1 + \Sigma_{+}) M \sin 2 \psi \right],
\end{split}\label{eq_eveqn_full_DE}\end{equation}
where
\begin{equation}
  Q = 2 \Sigma_{+}^{2} + R^{2} + \case{1}{6} \mathcal{H}^{2} + 
    \case{1}{2}(3 \gamma-2) \Omega , \label{eq_eveqn_Q}
\end{equation}
and
\begin{equation}
  \Omega = 1 - \Sigma_{+}^{2} - R^{2} - \Omega_{h} . \label{eq_eveqn_Omega}
\end{equation}
The evolution equation for $\Omega$ becomes
\begin{equation}
  \Omega ' = \left[ 2Q - (3\gamma -2) + 2 R^{2} \cos 2 \psi \right] \Omega .
    \label{eq_eveqn_Omega_DE}
\end{equation}
The restrictions \eref{eq_eveqn_restB} are equivalent to
\begin{equation}
  3 M^{2} R^{2} \sin^{2} \psi < 1, \qquad  M > 0, \qquad
  R \geq 0, \qquad \mathcal{H} > 0 . \label{eq_eveqn_BMVII0b}
\end{equation}


\section{Limits at late times} \label{sec_limits}

In this section we present a theorem which gives the limiting behaviour as
$\tau \rightarrow +\infty$ of the magnetic Bianchi VII$_{0}$
cosmologies when the equation of state parameter $\gamma$ satisfies
$\frac{2}{3} < \gamma < 2$. As a corollary of the theorem, we obtain the
limiting behaviour of certain dimensionless scalars that describe physical
properties of the models, namely the density parameter $\Omega$, defined by
\eref{eq_eveqn_Omega_defn}, the magnetic density parameter $\Omega_{h}$,
defined by \eref{eq_eveqn_Omegah_defn}, the \textit{shear parameter} $\Sigma$,
defined by
\begin{equation}
  \Sigma^{2} = \frac{ \sigma_{ab} \sigma^{ab} }{6H^{2}},
    \label{eq_limits_Sigma_defn}
\end{equation}
where $\sigma_{ab}$ is the rate-of-shear tensor of the fluid congruence, and
the \textit{Weyl curvature parameter} $\mathcal{W}$, defined by
\begin{equation}
  \mathcal{W}^{2} = \frac{ E_{ab} E^{ab} + H_{ab} H^{ab} }{6H^{4}},
    \label{eq_limits_Weyl_defn}
\end{equation}
where $E_{ab}$ and $H_{ab}$ are the electric and magnetic parts of the Weyl
tensor, respectively (see \cite{WE}, pg.~19), relative to the fluid congruence.

In terms of the Hubble-normalized variables, the shear parameter is given by
\begin{equation}
  \Sigma^{2} = \Sigma_{+}^{2} + R^{2} \cos^{2} \psi,
    \label{eq_limits_Sigma_explicit}
\end{equation}
which follows from \eref{eq_eveqn_cov} in conjunction with equation (6.13)
in \cite{WE}. The formula for the Weyl curvature parameter is more complicated
and is provided in appendix~C.

The main result concerning the limits of $\Sigma_{+}$, $R$, $\mathcal{H}$ and
$M$ is contained in the following theorem. Some of the results depend on
requiring that the model is not locally rotationally symmetric\footnote{See,
for example, \cite{WE}, pg.~22. We note that the LRS magnetic Bianchi
VII$_{0}$ models are described by the invariant subset $\Sigma_{-} = N_{-} =
0$, equivalently, $R = 0$. Since LRS models of Bianchi type VII$_{0}$ also
admit a group $G_{3}$ of isometries of Bianchi type I, we do not consider them
in detail here.} (LRS).
\begin{theorem}
  For all magnetic Bianchi cosmologies of type VII$_{0}$ that are
  not LRS and with density parameter $\Omega$ satisfying $\Omega > 0$, the
  Hubble-normalized state variables $(\Sigma_{+},R,\mathcal{H},M)$
  satisfy\footnote{The limits in the case $\gamma = \frac{4}{3}$ were
  conjectured by Sam Lisi. We thank him for helpful discussions.}
  \begin{equation}
    \lim_{\tau \rightarrow +\infty} (\Sigma_{+},R,\mathcal{H},M) = \begin{cases}
      (0,0,0,0), &\text{if}\quad \frac{2}{3} < \gamma < \frac{4}{3},\\
      \left(0, \sqrt{\case{2}{3}}\, k, \sqrt{2}\, k, 0\right), &\text{if}\quad
        \gamma = \frac{4}{3},\\
      \left(0, \sqrt{\case{2}{3}}, \sqrt{2}, 0\right), &\text{if}\quad
        \frac{4}{3} < \gamma < 2, \end{cases}
      \label{eq_limits_Hubblevars}
  \end{equation}
  and
  \begin{equation}
    \lim_{\tau \rightarrow +\infty} \frac{M}{R} = \begin{cases}
      +\infty, &\text{if}\quad \frac{2}{3} < \gamma < 1,\\
      L \not= 0, &\text{if}\quad \gamma = 1,\\
      0, &\text{if}\quad 1 < \gamma < 2, \end{cases} \label{eq_limits_MoverR}
  \end{equation}
  where $k \in (0,1)$ and $L > 0$ are constants that depend on the initial
  conditions. \label{theorem_limits}
\end{theorem}

\noindent \textbf{Proof.} It follows immediately from
\eref{eq_eveqn_N_plus_limit} and \eref{eq_eveqn_cov} that
\begin{equation}
  \lim_{\tau \rightarrow +\infty} M = 0 . \label{eq_limits_M}
\end{equation}
Furthermore, since $\Sigma_{+}$ is bounded, it follows from the $\psi$
evolution equation in \eref{eq_eveqn_full_DE} that
\begin{equation}
  \lim_{\tau \rightarrow +\infty} \psi = +\infty. \label{eq_limits_psi}
\end{equation}
The trigonometric functions in the DE~\eref{eq_eveqn_full_DE} thus oscillate
increasing rapidly as $\tau \rightarrow +\infty$. In order to control these
oscillations, we introduce new gravitational variables $\bar{\Sigma}_{+}$,
$\bar{R}$ and $\bar{\mathcal{H}}$ according to\footnote{We are motivated by
the analysis in \cite{HHTW03} (see equations~(3.8)) and \cite{WHU99} (see
equations~(B.4)).}
\begin{equation}\begin{split}
  \bar{\Sigma}_{+} &= \Sigma_{+} - \case{1}{4}(1 + \Sigma_{+})
    R^{2} M \sin 2 \psi , \\
  \bar{R} &= R \left[ 1 - \case{1}{4}(R^{2} - 1 - \Sigma_{+})
    M \sin 2 \psi \right] , \\
  \bar{\mathcal{H}} &= \mathcal{H} \left[ 1 - \case{1}{4} R^{2}
    M \sin 2 \psi \right].
\end{split}\label{eq_limits_cov}\end{equation}
These new variables are defined so as to `suppress' the rapidly oscillating
terms which may not tend to zero as $\tau \rightarrow +\infty$. The evolution
equations for these ``barred'' variables, which can be derived from
\eref{eq_eveqn_full_DE} and \eref{eq_limits_cov}, have the following form
\begin{equation}\begin{split}
  \bar{\Sigma}'_{+} &= -(2-\bar{Q}) \bar{\Sigma}_{+} - \bar{R}^{2} +
    \case{1}{3} \bar{\mathcal{H}}^{2} + M B_{\bar{\Sigma}_{+}} , \\
  \bar{R}' &= (\bar{Q} + \bar{\Sigma}_{+} - 1 + M B_{\bar{R}}) \bar{R}, \\
  \bar{\mathcal{H}}' &= (\bar{Q} - 2 \bar{\Sigma}_{+} - 1 +
    M B_{\bar{\mathcal{H}}}) \bar{\mathcal{H}} ,
\end{split}\label{eq_limits_barred_DE}\end{equation}
where
\begin{equation}
  \bar{Q} = 2 \bar{\Sigma}_{+}^{2} + \bar{R}^{2} + \case{1}{6} 
    \bar{\mathcal{H}}^{2} + \case{1}{2}(3\gamma-2) \left( 1 - 
    \bar{\Sigma}_{+}^{2} - \bar{R}^{2} - \case{1}{6} 
    \bar{\mathcal{H}}^{2} \right), \label{eq_limits_Q}
\end{equation}
and the $B$ terms are bounded functions in $\bar{\Sigma}_{+}$, $\bar{R}$,
$\bar{\mathcal{H}}$ and in $M$ and $\psi$ for $\tau$ sufficiently large.
The essential idea is to regard $M$ and $\psi$ as arbitrary functions of
$\tau$ subject only to \eref{eq_limits_M}. Thus, \eref{eq_limits_barred_DE}
is a non-autonomous DE for
$$
  \bar{\vec{x}} = ( \bar{\Sigma}_{+}, \bar{R}, \bar{\mathcal{H}} ),
$$
of the form
\begin{equation}
  \bar{\vec{x}}' = \vec{f}(\bar{\vec{x}}) + \vec{g}(\bar{\vec{x}},\tau),
    \label{eq_limits_nonautonomous_DE}
\end{equation}
where
\begin{equation}
  \vec{g}(\bar{\vec{x}},\tau) = M(\tau) ( B_{\bar{\Sigma}_{+}}, \bar{R}
    B_{\bar{R}}, \bar{\mathcal{H}} B_{\bar{\mathcal{H}}} ), \label{eq_limits_g}
\end{equation}
and $\vec{f}(\bar{\vec{x}})$ can be read off from the right-hand side of
\eref{eq_limits_barred_DE}. Since
$$
  \lim_{\tau \rightarrow +\infty} \vec{g}(\bar{\vec{x}},\tau) = \mathbf{0},
$$
as follows from \eref{eq_limits_M}, the DE~\eref{eq_limits_barred_DE} is
\textit{asymptotically autonomous} (see \cite{SY67}). The corresponding
autonomous DE is
\begin{equation}
  \hat{\vec{x}} = \vec{f}(\hat{\vec{x}}), \label{eq_limits_hatted_DE}
\end{equation}
where
$$
  \hat{\vec{x}} = ( \hat{\Sigma}_{+}, \hat{R}, \hat{\mathcal{H}} ).
$$

Using standard methods from the theory of dynamical systems, we first show that
the limits of the ``hatted'' variables correspond to those limits stated
in the theorem. Details are provided in appendix~B.1. We then use a theorem
from \cite{SY67} (see theorem~\ref{theorem_SY} in appendix~B) to infer that
the solutions of the non-autonomous DE~\eref{eq_limits_nonautonomous_DE} have
the same limits as the solutions of the autonomous
DE~\eref{eq_limits_hatted_DE}. Details are provided in appendix~B.2. The
limit of $\vec{x} = (\Sigma_{+},R,\mathcal{H})$ follows immediately from this
result in conjunction with the definitions~\eref{eq_limits_cov}. Finally, the
limit~\eref{eq_limits_MoverR} concerning the ratio $M/R$ is obvious when
$\frac{4}{3} \leq \gamma < 2$, since $R \nrightarrow 0$. The more complicated
case when $\frac{2}{3} < \gamma < \frac{4}{3}$ is treated in
appendix~B.3.\hfill $\Box$

\begin{table}[t] \renewcommand{\arraystretch}{1.3}
\begin{minipage}{\linewidth}
\renewcommand{\thefootnote}{\thempfootnote}
\renewcommand{\thempfootnote}{\fnsymbol{mpfootnote}}
\renewcommand{\footnoterule}{\vspace{0ex}}
\begin{center} \small
\caption{\label{table_limits_scalars}Limits of the Hubble-normalized scalars
  $\Omega$, $\Omega_{h}$, $\Sigma$ and $\mathcal{W}$ at late times for
  magnetic Bianchi cosmologies of type VII$_{0}$.}\vspace{1.5ex}
\begin{tabular}{lllll} \hline\hline
  \multicolumn{1}{l}{\rule[-2mm]{0mm}{6.5mm}Range of $\gamma$} &  $\Omega$ &
    $\Omega_{h}$ & $\Sigma^{2}$ & $\mathcal{W}$ \\ \hline
  $\frac{2}{3} < \gamma < 1$ & 1 & 0 & 0 & 0 \\
  $\gamma = 1$ & 1 & 0 & 0 & $L \not= 0$ \\
  $1 < \gamma < \frac{4}{3}$ & 1 & 0 & 0 & $+\infty$ \\
  $\gamma = \frac{4}{3}$ & $1 - k^{2}$ & $\frac{1}{3} k^{2}$ &
    $\left(0,\frac{2}{3} k^{2}\right)$\footnote[2]{The components in the
    parentheses are the $\liminf$ and $\limsup$. The parameter $k$ is the
    parameter that appears in theorem~\ref{theorem_limits}.} & $+\infty$ \\
  $\frac{4}{3} < \gamma < 2$ & 0 & $\frac{1}{3}$ & $\left(0,\frac{2}{3}\right)$
    & $+\infty$ \\ \hline\hline
\end{tabular}\end{center}\end{minipage}
\end{table}

\begin{corollary}
  The limits as $\tau \rightarrow +\infty$ of the density parameter $\Omega$,
  the magnetic density parameter $\Omega_{h}$, the shear scalar $\Sigma$ and
  the Weyl curvature scalar $\mathcal{W}$, for all magnetic
  Bianchi cosmologies of type VII$_{0}$ that are not LRS and with $\Omega$
  satisfying $\Omega > 0$, are as given in table~\ref{table_limits_scalars}.
  \label{cor_limits_scalars}
\end{corollary}

\noindent \textbf{Proof.} These results follow directly from
theorem~\ref{theorem_limits} and equations \eref{eq_eveqn_Omega},
\eref{eq_limits_Sigma_explicit}, \eref{eq_app_Weyl_c} and
\eref{eq_app_Weyl_explicit}. Moreover, if $1 < \gamma < 2$, it follows from
\eref{eq_app_Weyl_c} and \eref{eq_app_Weyl_explicit} that since $\Sigma_{+}$,
$R$ and $\mathcal{H}$ are bounded, and $\lim_{\tau \rightarrow +\infty} M/R =
0$, that
$$
  \mathcal{W} = \frac{2R}{M} [1 + \bigO(M)],
$$
as $\tau \rightarrow +\infty$.\hfill $\Box$

We conclude this section by discussing the physical interpretations of
theorem~\ref{theorem_limits} and its corollary. Like the non-magnetic Bianchi
VII$_{0}$ models, the magnetic Bianchi VII$_{0}$ models are not asymptotically
self-similar at late times, since the orbits in the Hubble-normalized state
space do not approach an equilibrium point of the evolution equations. This
phenomenon is accompanied by Weyl curvature dominance, characterized by the
divergence of the Weyl curvature scalar $\mathcal{W}$ which describes the
intrinsic anisotropy of the gravitational field. For non-LRS models with
$1 < \gamma < 2$, $\mathcal{W}$ is unbounded as $\tau \rightarrow +\infty$.

The shear scalar $\Sigma$ quantifies the anisotropy in the expansion of a
cosmological model. We see that for $\frac{2}{3} < \gamma < \frac{4}{3}$, the
models isotropize at late times in the sense that $\lim_{\tau \rightarrow
+\infty} \Sigma = 0$, as is the case for the corresponding non-magnetic Bianchi
VII$_{0}$ models (see \cite{WHU99}, theorem~2.3). The key difference between
the magnetic and non-magnetic models occurs when the matter content is a
radiation fluid ($\gamma = \frac{4}{3}$): the presence of a magnetic field
prevents shear isotropization, in the sense that $\Sigma$ does not tend to
zero at late times.


\section{The singular asymptotic regime} \label{sec_past}

In this section we show, by combining numerical experiments with analytical
considerations, that generic non-LRS magnetic Bianchi VII$_{0}$ cosmologies
with $\frac{2}{3} < \gamma < 2$ exhibit an oscillatory approach to the initial
singularity, as do the other class~A magnetic Bianchi models.

\begin{figure}[t] \small
  \begin{center}\includegraphics[scale=0.55]{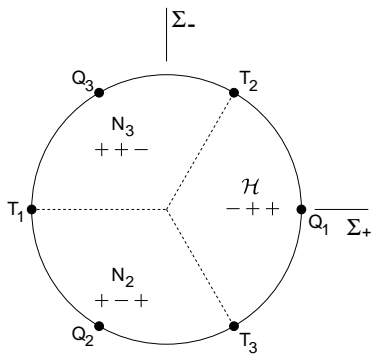}\end{center}
  \caption{\label{fig_past_Kasner}The arrays $-++$, etc.\ give the signs of
    the eigenvalues $\lambda_{\mathcal{H}}$, $\lambda_{N_{2}}$ and
    $\lambda_{N_{3}}$ in that order. The variables listed next to each of the
    three arcs indicates which of the variables $\mathcal{H}$, $N_{2,3}$ is
    growing into the past.}
\end{figure}

As is the case for the previously studied class~A magnetic Bianchi models, the
behaviour into the past for the magnetic Bianchi VII$_{0}$ models is
necessarily complicated since none of the equilibrium points of the evolution
equations~\eref{eq_eveqn_DE_nonpolar} are local sources. It is well-known that
in the dynamical systems approach, the Kasner circle $\mathcal{K}$ plays a
primary role in determining the dynamics towards the singularity, since its
local stability enables one to predict whether the singularity in a given
class of models is oscillatory\footnote{In a recent paper \cite{UEWE03}, it has
been shown that the Kasner circle also plays this role in cosmological models
without symmetry.}. For the present class of models, $\mathcal{K}$ is the
set of equilibrium points described by\footnote{In discussing the singular
asymptotic regime, it is more convenient to use the spatial connection
variables $N_{2}$ and $N_{3}$, rather than $N_{+}$ and $N_{-}$ (see
equation~\eref{eq_eveqn_Npm}).}
$$
  \Sigma_{+}^{2} + \Sigma_{-}^{2} = 1, \qquad N_{2} = N_{3} = \mathcal{H} = 0 .
$$
A local stability analysis shows that the Kasner equilibrium points are saddles
in the Hubble-normalized state space. Apart from three exceptional points
(labeled $T_{1}$, $T_{2}$ and $T_{3}$ in figure~\ref{fig_past_Kasner}), the
equilibrium points of $\mathcal{K}$ have a one-dimensional unstable manifold
into the past. Figure~\ref{fig_past_Kasner} shows the signs of the eigenvalues
on $\mathcal{K}$ associated with the variables $\mathcal{H}$, $N_{2}$ and
$N_{3}$ (a negative eigenvalue indicates instability into the past) and which
of these three variables are increasing into the past in a neighbourhood of
the Kasner circle.

\begin{figure}[t] \small
  \begin{tabular}{ccc}
    \includegraphics[scale=0.46]{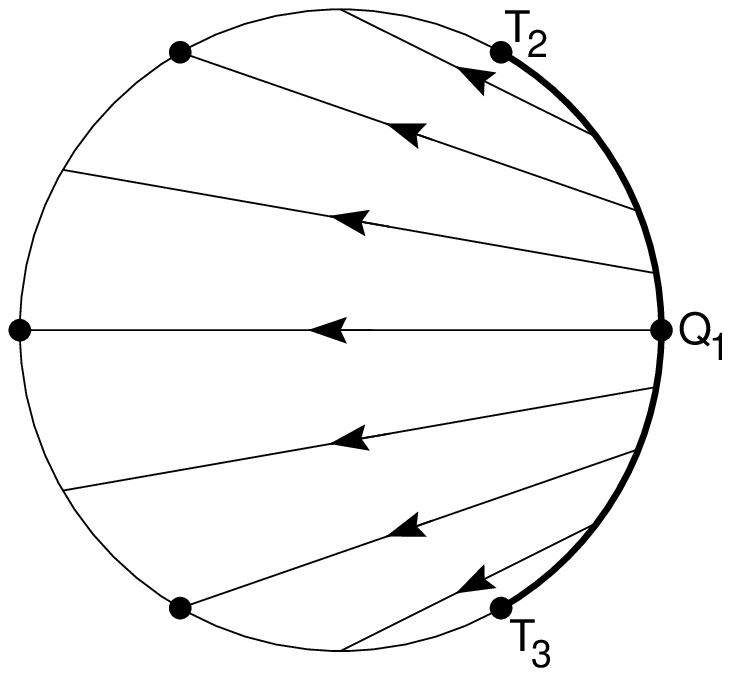} &
    \includegraphics[scale=0.46]{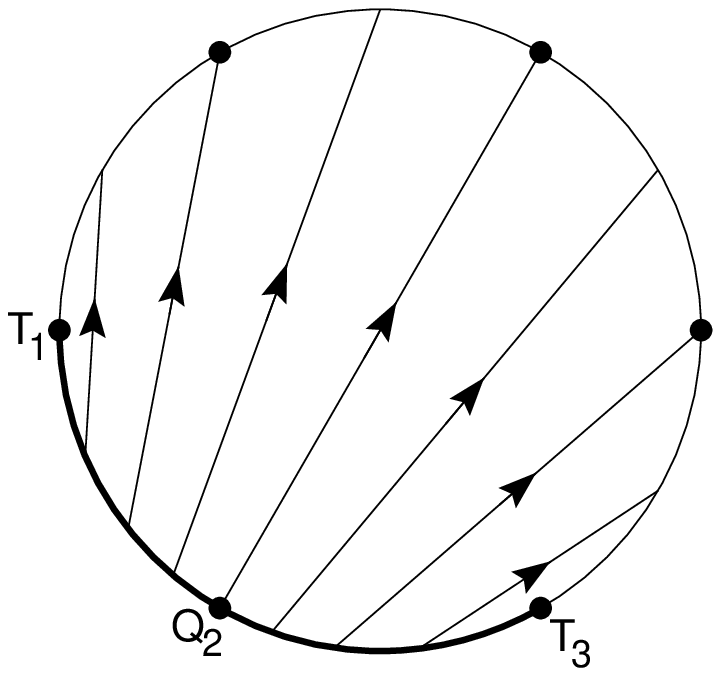} &
    \includegraphics[scale=0.46]{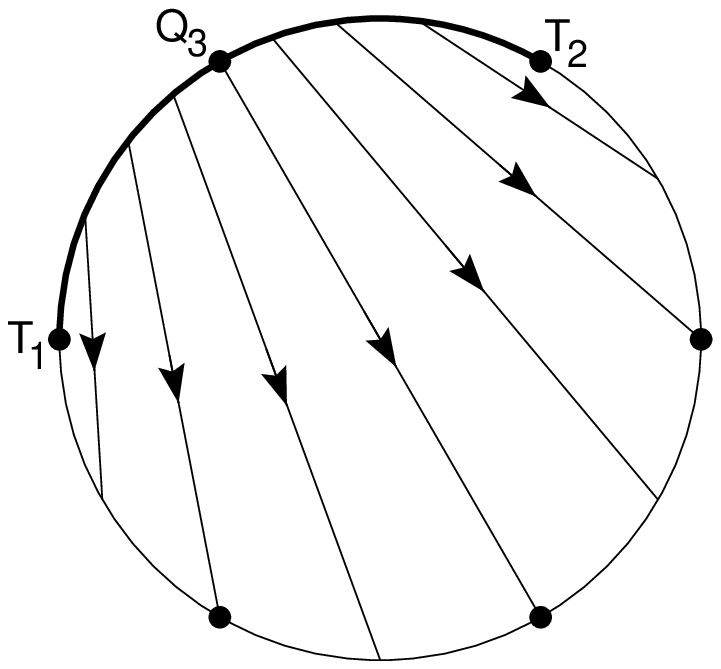} \\
    \small $\mathcal{S}_{\mathcal{H}}$ &
    \small $\mathcal{S}_{N_{2}}$ &
    \small $\mathcal{S}_{N_{3}}$
  \end{tabular}
  \caption{\label{fig_past_S}The projections of the Rosen and Taub orbits
    joining points on the Kasner circle $\mathcal{K}$. The arrows show
    evolution into the past.}
\end{figure}

It turns out each unstable manifold on $\mathcal{K}$ is asymptotic to another
Kasner point. In other words, the unstable manifold is a \textit{heteroclinic
orbit} of $\mathcal{K}$, i.e.\ an orbit which joins two Kasner points. These
unstable manifolds provide a mechanism for a cosmological model to make a
transition from one (approximate) Kasner state to another, as it evolves
into the past. Figure~\ref{fig_past_S} shows the projections in the
$\Sigma_{+}\Sigma_{-}$-plane, of these families of heteroclinic orbits, that
join two Kasner points. The families are described as follows:
\begin{equation}
  \renewcommand{\arraystretch}{1.3}
  \begin{array}{llll}
    \mathcal{S}_{\mathcal{H}} : &\quad \Sigma_{+}^{2} + \Sigma_{-}^{2} +
      \case{1}{6} \mathcal{H}^{2} = 1, &\quad \mathcal{H} > 0, &\quad
      N_{2} = N_{3} = 0, \\
    \mathcal{S}_{N_{2}} : &\quad \Sigma_{+}^{2} + \Sigma_{-}^{2} + \case{1}{12}
      N_{2}^{2} = 1, &\quad N_{2} > 0, &\quad N_{3} = \mathcal{H} = 0, \\
    \mathcal{S}_{N_{3}} : &\quad \Sigma_{+}^{2} + \Sigma_{-}^{2} + \case{1}{12}
      N_{3}^{2} = 1, &\quad N_{3} > 0, &\quad N_{2} = \mathcal{H} = 0 .
  \end{array} \label{eq_past_S}
\end{equation}
The heteroclinic orbits on $\mathcal{S}_{N_{2,3}}$ describe the familiar
vacuum Bianchi II Taub models, while the orbits on $\mathcal{S}_{\mathcal{H}}$
describe the Rosen magneto-vacuum models (see \cite{LKW95}, pg.~531).

Numerical experiments suggest that for generic non-LRS models, after an
initial transient stage, the orbit approaches a point on $\mathcal{K}$. The
direction of departure of the orbit is determined by the unique Rosen or
Taub orbit through that point whereupon it shadows (i.e.\ is approximated by)
this orbit until is approaches another point on $\mathcal{K}$ and the
process repeats indefinitely. In physical terms, the corresponding 
cosmological model is approximated by an infinite sequence of Kasner vacuum
models as the singularity is approached into the past, the so-called
\textit{Mixmaster oscillatory singularity}. This behaviour motivates the
following conjecture concerning the past attractor $\mathcal{A}^{-}$.

\begin{conjecture}
  The past attractor is the two-dimensional invariant set consisting of all
  orbits in the invariant sets $\mathcal{S}_{\mathcal{H}}$,
  $\mathcal{S}_{N_{2}}$ and $\mathcal{S}_{N_{3}}$ (see figure~\ref{fig_past_S})
  and the Kasner equilibrium points, i.e.
  \begin{equation}
    \mathcal{A}^{-} = \mathcal{S}_{\mathcal{H}} \cup \mathcal{S}_{N_{2}}
      \cup \mathcal{S}_{N_{3}} \cup \mathcal{K} .
      \label{eq_past_pastattractor}
  \end{equation} \label{conjecture_past}
\end{conjecture}

This conjecture can be formulated in terms of limits of the state variables
as follows. Referring to \eref{eq_eveqn_Omega_nonpolar} and \eref{eq_past_S},
we see that the set $\mathcal{A}^{-}$ is defined by $\Omega = 0$ and
$$
  N_{2} \mathcal{H} = N_{3} \mathcal{H} = N_{2} N_{3} = 0 .
$$
It follows from monotone function arguments (see the comment at the end of
appendix~A) that
$$
  \lim_{\tau \rightarrow -\infty} N_{2} N_{3} = 0,
$$
and, moreover, that $N_{2}$ and $N_{3}$ {are bounded in the singular regime}.
Thus our conjecture concerning the past attractor can be formulated as
$$
  \lim_{\tau \rightarrow -\infty} \Omega = 0, \qquad
  \lim_{\tau \rightarrow -\infty} N_{2} \mathcal{H} = 0 , \qquad
  \lim_{\tau \rightarrow -\infty} N_{3} \mathcal{H} = 0 .
$$
Note that for a generic orbit, $\lim_{\tau \rightarrow -\infty}
(\mathcal{H},N_{2},N_{3})$ does not exist.


\section{Discussion} \label{sec_discussion}

With the appearance of the present paper there is now available a complete
description of the dynamics of magnetic Bianchi cosmologies\footnote{We
emphasize that we are restricting our attention to Bianchi cosmologies that are
non-tilted, in the sense that the fluid four-velocity is orthogonal to the
group orbits.} with a perfect-fluid matter content, in the two asymptotic
regimes. We now give an overview of the properties of these models, in order to
highlight the role of a primordial magnetic field in spatially homogeneous
cosmological dynamics. The possible Bianchi types and relevant references are
given in the introduction. We emphasize the so-called class~A models (in the
terminology of Ellis and MacCallum \cite{EM69}), that is, those of Bianchi
types I, II, VI$_{0}$ and VII$_{0}$. For each of these types the
Hubble-normalized state space is five-dimensional, but, as indicated in
table~\ref{table_dof}, they differ as regards the number of degrees of freedom
associated with spatial curvature and with the magnetic field. In this table we
have also listed Bianchi types VIII and IX, which do not admit a magnetic
field, for comparison purposes.

\begin{table}[t] \renewcommand{\arraystretch}{1.3}
\begin{center} \small
\caption{\label{table_dof}Shear, spatial curvature and magnetic degrees of
    freedom in class~A Bianchi cosmologies.}\vspace{1.5ex}
\begin{tabular}{llll} \hline\hline
  \multicolumn{1}{l}{\rule[-2mm]{0mm}{6.5mm}Bianchi type} & Shear &
    Spatial curvature & Magnetic field \\ \hline
  I & 2 & 0 & 3 \\
  II & 2 & 1 & 2 \\
  VI$_{0}$, VII$_{0}$ & 2 & 2 & 1 \\
  VIII, IX & 2 & 3 & 0 \\ \hline\hline
\end{tabular}\end{center}
\end{table}

All models in table~\ref{table_dof} display an oscillatory approach to the
singularity, described by a two-dimensional attractor in the Hubble-normalized
state space, familiar from the non-magnetic Bianchi VIII and IX models
(see \cite{WE}, pp.~143--7). The essential point is that the magnetic field
mimics spatial curvature in that it destabilizes the Kasner circle of
equilibrium points. Note that the sum of the number of spatial curvature
and magnetic degrees of freedom is three, in all cases in
table~\ref{table_dof}.

\begin{table}[t] \renewcommand{\arraystretch}{1.3}
\begin{minipage}{\linewidth}
\renewcommand{\thefootnote}{\thempfootnote}
\renewcommand{\thempfootnote}{\fnsymbol{mpfootnote}}
\renewcommand{\footnoterule}{\vspace{0ex}}
\begin{center} \small
\caption{\label{table_dust}Limiting values of $\Omega$, $\Omega_{h}$, $\Sigma$
and $\mathcal{W}$ as $\tau \rightarrow +\infty$ for magnetic Bianchi
cosmologies with a dust fluid ($\gamma = 1$).}\vspace{1.5ex}
\begin{tabular}{lllll} \hline\hline
  \multicolumn{1}{l}{\rule[-2mm]{0mm}{6.5mm}Bianchi type} &  $\Omega$ &
    $\Omega_{h}$ & $\Sigma^{2}$ & $\mathcal{W}^{2}$ \\ \hline
  I & 1 & 0 & 0 & 0 \\
  II & $\frac{15}{16}$ & 0 & $\frac{1}{64}$ & $\frac{45}{2048}$ \\
  VI$_{0}$\footnote[2]{The parameter $k$ satisfies $0 < k < 1$ and depends on
    the initial conditions.} & $\frac{3}{4}(1-k^{2})$ & $\frac{3}{8} k^{2}$ &
    $\frac{1}{16}$ & $\frac{9}{128}(1+2k^{2})(2+k^{2})$ \\
  VII$_{0}$ & 1 & 0 & 0 & $L > 0$ \\ \hline\hline
\end{tabular}\end{center}\end{minipage}
\end{table}

As regards the late-time dynamics of the magnetic cosmologies, there is a
fundamental difference between the Bianchi VII$_{0}$ models considered in
the present paper and the Bianchi I, II and VI$_{0}$ models considered
earlier (\cite{LKW95}, \cite{L97}, \cite{L98}), as follows. The Bianchi I, II
and VI$_{0}$ models are asymptotically self-similar, in the sense that each
model is approximated by an exact self-similar solution, while the models
of type VII$_{0}$ are not asymptotically self-similar. This difference is
essentially a consequence of the fact that the Hubble-normalized state space
of the Bianchi VII$_{0}$ models is unbounded. Another feature of the magnetic
cosmologies is that the asymptotic dynamics at late times depends
significantly on the equation of state parameter $\gamma$. We illustrate this
dependence in tables~\ref{table_dust} and \ref{table_radiation} where we
give the limits at late times of $\Omega$, $\Omega_{h}$, $\Sigma$ and
$\mathcal{W}$ for the two physically important cases, dust ($\gamma = 1$) and
radiation ($\gamma = \frac{4}{3}$). It is worthy of note that if $\gamma <
\frac{4}{3}$, the Bianchi I models isotropize in all respects ($\Sigma
\rightarrow 0$, $\mathcal{W} \rightarrow 0$ and $\Omega_{h} \rightarrow 0$)
while the Bianchi VII$_{0}$ models isotropize as regards the shear and the
magnetic field ($\Sigma \rightarrow 0$ and $\Omega_{h} \rightarrow 0$).

A cosmic magnetic field also affects the local stability of the equilibrium
point that corresponds to the flat FL solution\footnote{In the present paper,
this equilibrium point is given by $\Sigma_{\pm} = 0$, $N_{\pm} = 0$,
$\mathcal{H} = 0$.}. For non-magnetic models, the flat FL equilibrium point
is typically a saddle point, having both a non-trivial stable manifold and
a non-trivial unstable manifold. The shear degrees of freedom generate the
stable manifold while the spatial curvature degrees of freedom generate the
unstable manifold. The stable manifold leads to the phenomenon of
\textit{intermediate isotropization}, i.e.\ a model can evolve to become
arbitrarily close to isotropy over a finite interval of time. The unstable
manifold leads to models with an \textit{isotropic singularity}, i.e.\ models
which are highly isotropic near the initial singularity, but which subsequently
develop anisotropies. The effect of a primordial magnetic field on these
phenomena depends on the equation of state parameter $\gamma$. If $\gamma <
\frac{4}{3}$, the magnetic field increases the dimension of the stable
manifold, leaving the dimension of the unstable manifold unchanged, thus
increasing the likelihood of intermediate isotropization. On the other hand,
if $\gamma > \frac{4}{3}$, the magnetic field increases the dimension of the
unstable manifold by one, leading to magnetic models with an isotropic
singularity.

\begin{table}[t] \renewcommand{\arraystretch}{1.3}
\begin{minipage}{\linewidth}
\renewcommand{\thefootnote}{\thempfootnote}
\renewcommand{\thempfootnote}{\fnsymbol{mpfootnote}}
\renewcommand{\footnoterule}{\vspace{0ex}}
\begin{center} \small
\caption{\label{table_radiation}Limiting values of $\Omega$, $\Omega_{h}$,
$\Sigma$ and $\mathcal{W}$ as $\tau \rightarrow +\infty$ for magnetic Bianchi
cosmologies with a radiation fluid ($\gamma = \frac{4}{3}$).}\vspace{1.5ex}
\begin{tabular}{lllll} \hline\hline
  \multicolumn{1}{l}{\rule[-2mm]{0mm}{6.5mm}Bianchi type} &  $\Omega$ &
    $\Omega_{h}$ & $\Sigma^{2}$ & $\mathcal{W}^{2}$ \\ \hline
  I & 1 & 0 & 0 & 0 \\
  II & $\frac{3}{4}$ & $\frac{1}{12}$ & $\frac{1}{12}$ & $\frac{21}{144}$ \\
  VI$_{0}$ & 0 & $\frac{3}{8}$ & $\frac{1}{16}$ & $\frac{81}{128}$ \\
  VII$_{0}$\footnote[2]{The parameter $k$ satisfies $0 < k < 1$ and depends on
    the initial conditions. The components in the parentheses for $\Sigma^{2}$
    correspond to its $\liminf$ and $\limsup$.} & $1 - k^{2}$ & $\frac{1}{3}
    k^{2}$ & $\left(0,\frac{2}{3} k^{2}\right)$ & $+\infty$ \\ \hline\hline
\end{tabular}\end{center}\end{minipage}
\end{table}

We conclude by giving some suggestions for future research. Firstly, it would
be of interest to investigate the asymptotic dynamics of \textit{spatially
inhomogeneous} cosmological models in the presence of a primordial magnetic
field, in order to determine which features of magnetic Bianchi cosmologies
occur in models without symmetries. The recent paper \cite{UEWE03} on $G_{0}$
cosmologies would provide a suitable framework for such an investigation. One
step has been taken in this direction by Weaver \etal \cite{WIB98}, who
investigated a family of inhomogeneous cosmologies that generalize the
magnetic Bianchi VI$_{0}$ cosmologies, and provided numerical evidence that the
singularity is oscillatory.

Secondly, the work of Barrow \etal \cite{B97} referred to in the
introduction leads to an upper bound on the magnetic density parameter
$\Omega_{h}$ of order $10^{-10}$ for spatially homogeneous magnetic fields.
It would be of interest to what extent this upper bound would be weakened
within the class of spatially inhomogeneous magnetic cosmologies. The
analysis of the anisotropies in the CMBR by Maartens \etal \cite{MES96} would
probably provide a suitable framework for such an analysis.

We finally comment briefly on other recent work on magnetic fields in
cosmology, which has focused on the potential dynamical effects of a
primordial magnetic field in a perturbed FL cosmology (\cite{MTU01},
\cite{MT00}, \cite{T01}, \cite{TB97}, \cite{TB98}, \cite{TM00b}), or in a
perturbed Bianchi~I cosmology (\cite{TM00a}). This work, which makes use of the
Ellis-Bruni covariant and gauge-invariant method for analyzing cosmological
density perturbations, complements the results in our paper and related ones,
which focus on the dynamics in the asymptotic regimes of magnetic Bianchi
cosmologies, and on the likelihood that such a model will evolve to be close
to FL. Extending the dynamical systems analysis of magnetic Bianchi
cosmologies to spatially inhomogeneous magnetic cosmologies may help to bridge
the gap between these two bodies of work.


\section*{Acknowledgements}

This research was funded by the Natural Sciences and Engineering Research
Council of Canada through a discovery grant (JW) and an Undergraduate Research
Award to JH (in 2002).


\appendix
\def\thesection{\Alph{section}}

\stepcounter{section}
\section*{Appendix~\thesection}

In this appendix we prove \eref{eq_eveqn_N_plus_limit}, concerning the limit
of the Hubble-normalized variable $N_{+}$ in the late-time regime. This result
is restated as proposition~\ref{prop_Nplus} below.

\begin{proposition}
  For all magnetic Bianchi cosmologies of type VII$_{0}$ that are not
  LRS\footnote{The result also holds for the LRS models. The proof is similar to
  the non-LRS case; we omit the details in this paper.}, with equation
  of state parameter $\gamma$ subject to $\frac{2}{3} < \gamma < 2$, and
  density parameter $\Omega$ satisfying $\Omega > 0$, the Hubble-normalized
  variable $N_{+}$ satisfies
  \begin{equation}
    \lim_{\tau \rightarrow +\infty} N_{+} = +\infty .
      \label{eq_app_Nplus_limit}
  \end{equation} \label{prop_Nplus}
\end{proposition}

\noindent{\textbf{Proof.}} The proof is similar to the proof of the
corresponding result for non-magnetic Bianchi VII$_{0}$ cosmologies (see
theorem~2.1 and equation~(A.5) in \cite{WHU99}), in that it makes use of
monotone functions and the so-called monotonicity principle (see chapter~4
in \cite{WE}). There are two cases depending on the value of $\gamma$.

\vspace{1.5ex plus 0.3ex minus 0.3ex}
\noindent \textit{Case 1. $\frac{2}{3} < \gamma \leq 1$}
\vspace{1.5ex plus 0.3ex minus 0.3ex}

\noindent As in \cite{WHU99}, we consider the function
$$
  Z_{1} = \frac{ (N_{+}^{2} - 3 N_{-}^{2})^{v} \Omega }{ (1 + v
    \Sigma_{+})^{2(1+v)} },
$$
with $v = \frac{1}{4}(3\gamma - 2)$. The evolution equations
\eref{eq_eveqn_DE_nonpolar} imply that
$$
  \frac{Z'_{1}}{Z_{1}} = \frac{4[(\Sigma_{+}+v)^{2}+(1-v^{2})\Sigma_{-}^{2}]}
    {1+v\Sigma_{+}} + \frac{(1+v)(1-4v)\mathcal{H}^{2}}{3(1 + v \Sigma_{+})}.
$$
Since $0 < v \leq \frac{1}{4}$ in this case, it follows that $Z_{1}$ is
monotone increasing and the result \eref{eq_app_Nplus_limit} follows as in
the non-magnetic case (see appendix~A in \cite{WHU99}).

\vspace{1.5ex plus 0.3ex minus 0.3ex}
\noindent \textit{Case 2. $1 < \gamma < 2$}
\vspace{1.5ex plus 0.3ex minus 0.3ex}

\noindent We give a proof by contradiction. Suppose that
\eref{eq_app_Nplus_limit} does not hold. Since the remaining variables are
bounded, it follows that for any point $\vec{x}$ in the state space the
$\omega$-limit set $\omega(\vec{x})$ is non-empty.

Consider the function
\begin{equation}
  Z_{2} = \frac{\Sigma_{-}^{2} + N_{-}^{2}}{N_{+}^{2} - 3 N_{-}^{2}},
    \label{eq_app_Nplus_B}
\end{equation}
which satisfies $0 < Z_{2} < +\infty$ on the invariant set $S$ defined by
\begin{equation}
  N_{+} > 0, \qquad N_{+}^{2} - 3 N_{-}^{2} > 0, \qquad
  \Sigma_{-}^{2} + N_{-}^{2} > 0, \qquad \Omega \geq 0, \qquad
  \mathcal{H} \geq 0. \label{eq_app_Nplus_S}
\end{equation}
The evolution equations \eref{eq_eveqn_DE_nonpolar} imply that
$$
  \frac{Z'_{2}}{Z_{2}} = \frac{-4(1 + \Sigma_{+})\Sigma_{-}^{2}}{\Sigma_{-}^{2}
    + N_{-}^{2}}.
$$
It follows that $Z_{2}$ is decreasing\footnote{No orbit in $S$ satisfies
$\Sigma_{+} =-1$ for all $\tau$.} along orbits in $S$. We can now apply the
monotonicity principle. By (A.3) the set $\bar{S} \setminus S$ (the set of
boundary points of $S$ that are not contained in $S$) is defined by one or
both of the following equalities holding
\begin{equation}
  N_{+}^{2} - 3 N_{-}^{2} = 0, \qquad \Sigma_{-}^{2} + N_{-}^{2} = 0.
    \label{eq_app_Nplus_bS}
\end{equation}
It now follows that for any $\vec{x} \in S$, the $\omega$-limit set
$\omega(\vec{x})$ is contained in the subset of $\bar{S} \setminus S$ that
satisfies $\lim_{\vec{y} \rightarrow \vec{s}} Z_{2}(\vec{y}) \not= +\infty$,
where $\vec{s} \in \bar{S} \setminus S$ and $\vec{y} \in S$. On account of
\eref{eq_app_Nplus_B} and \eref{eq_app_Nplus_bS} we conclude that
\begin{equation}
  \omega(\vec{x}) \subset \{ \vec{x} \:|\: \Sigma_{-} = N_{-} = 0 \}.
    \label{eq_app_Nplus_C}
\end{equation}
We can further restrict the possible $\omega$-limit sets by considering the
function
$$
  Z_{3} = \frac{\Omega^{2}}{(N_{+}^{2} - 3 N_{-}^{2}) \mathcal{H}^{2}},
$$
which satisfies
$$
  Z'_{3} = -6 (\gamma - 1) Z_{3},
$$
as follows from \eref{eq_eveqn_DE_nonpolar}. We immediately conclude that
$\lim_{\tau \rightarrow +\infty} Z_{3} = 0$ and hence that $\lim_{\tau
\rightarrow +\infty} \Omega = 0$.  In conjunction with \eref{eq_app_Nplus_C},
this result implies that $\omega(\vec{x}) \subset S_{1}$, where
$$
  S_{1} = \{ \vec{x} \:|\: \Sigma_{-} = N_{-} = \Omega = 0 \} .
$$
The only potential $\omega$-limit sets in $S_{1}$ are equilibrium points,
since\footnote{On $S_{1}$ the evolution equation for $N_{+}$ reduces to
$N'_{+} = (1 + \Sigma_{+})^{2} N_{+}$.} $\lim_{\tau \rightarrow +\infty}
N_{+} = +\infty$ for all other orbits in $S_{1}$. The equilibrium points are
\begin{itemize}
  \item[i.] $\Sigma_{+} = 1, \quad N_{+} = \mathcal{H} = 0, \quad
    \text{(an isolated point)}$\vspace{-1ex plus 0.2ex minus 0.2ex}
  \item[ii.] $\Sigma_{+} = -1, \quad \mathcal{H} = 0, \quad N_{+} > 0 \quad
    \text{(a line)}$.
\end{itemize}
No orbit with $\mathcal{H} > 0$, $\Omega > 0$ and $1 < \gamma < 2$ can be
future asymptotic to any of these equilibrium points, since $\Omega' > 0$ in
a neighbourhood of any of these points as follows from
\eref{eq_eveqn_Omega_DE_nonpolar} and \eref{eq_eveqn_q_nonpolar}. Thus we
have a contradiction of the fact that $\omega(\vec{x}) \not= \phi$, and as a
result, \eref{eq_app_Nplus_limit} holds in
case~2.\hfill $\Box$

\vspace{1.5ex plus 0.3ex minus 0.3ex}
\noindent \textbf{Comment.} The monotone function \eref{eq_app_Nplus_B} also
provides useful information about the \textit{past asymptotics} of magnetic
Bianchi cosmologies. From the monotonicity principle, we can conclude that
$$
  \alpha(\vec{x}) \subset \{ \vec{x} \:|\: N_{+}^{2} - 3 N_{-}^{2} = 0 \},
$$
for any $\vec{x} \in S$. Therefore, in contrast to the late-time regime,
\textit{$N_{+}$ is bounded towards the initial singularity.}


\stepcounter{section}
\section*{Appendix~\thesection}

In this appendix we fill in the details of the proof of
theorem~\ref{theorem_limits}. The proof of this theorem relies on a result of
\cite{SY67} (see corollary 3.3, pg.~180) concerning asymptotically
autonomous DEs, stated as theorem~\ref{theorem_SY} below.

Consider a non-autonomous DE
\begin{equation}
  \bar{\vec{x}}' = \vec{f}(\bar{\vec{x}}) + \vec{g}(\bar{\vec{x}},\tau),
    \label{eq_app_SY_nonautonomous_DE}
\end{equation}
and the associated autonomous DE
\begin{equation}
  \hat{\vec{x}}' = \vec{f}(\hat{\vec{x}}), \label{eq_app_SY_autonomousDE}
\end{equation}
where $\vec{f} : D \rightarrow \mathbb{R}^{n}$, $\vec{g} : D \times
\mathbb{R} \rightarrow \mathbb{R}^{n}$ and $D$ is an open subset of
$\mathbb{R}^{n}$. It is assumed that
\newcounter{mycounter}
\begin{list}{\arabic{mycounter}\hfill}
  {\usecounter{mycounter}
   \setlength{\leftmargin}{10mm}
   \setlength{\labelwidth}{10mm}
   \setlength{\labelsep}{0mm}
   \setlength{\itemsep}{-0.5ex plus 0.2ex minus 0.2ex}
   \setlength{\topsep}{1ex plus 0.1ex minus 0.1ex} }
  \item[$H_{1}$:\hfill] $\displaystyle \lim_{\tau \rightarrow +\infty}
    \vec{g}(\vec{w}(\tau),\tau) = \mathbf{0}$ for every continuous
    function $\vec{w} : [\tau_{0},+\infty) \rightarrow D$
  \item[and\hfill] \hspace{1em}
  \item[$H_{2}$:\hfill] any solution of \eref{eq_app_SY_nonautonomous_DE}
    with initial condition in $D$ is bounded for $\tau \geq \tau_{0}$, for
    some $\tau_{0}$ sufficiently large.
\end{list}

\begin{theorem}
  If $H_{1}$ and $H_{2}$ are satisfied and any solution of
  \eref{eq_app_SY_autonomousDE} with initial condition in $D$ satisfies
  $$
    \lim_{\tau \rightarrow +\infty} \hat{\vec{x}}(\tau) = \vec{a},
  $$
  then any solution of \eref{eq_app_SY_nonautonomous_DE} with initial
  condition in $D$ satisfies
  $$
    \lim_{\tau \rightarrow +\infty} \bar{\vec{x}}(\tau) = \vec{a}.
  $$ \label{theorem_SY}
\end{theorem} 

\noindent We make use of this theorem in appendix~B.2.


\stepcounter{subsection}
\subsection*{Appendix~\thesubsection}

We now deduce the limits at late times of $(\hat{\Sigma}_{+},\hat{R},
\hat{\mathcal{H}})$. The components of the DE~\eref{eq_limits_hatted_DE},
$\hat{\vec{x}}' = \vec{f}(\hat{\vec{x}})$, are given by
\begin{equation}\begin{split}
  \hat{\Sigma}'_{+} &= (\hat{Q}-2)\hat{\Sigma}_{+} - \hat{R}^{2} + 
    \case{1}{3} \hat{\mathcal{H}}^{2}, \\
  \hat{R}' &= (\hat{Q}+\hat{\Sigma}_{+}-1) \hat{R}, \\
  \hat{\mathcal{H}}' &= (\hat{Q}-2\hat{\Sigma}_{+}-1) \hat{\mathcal{H}},
\end{split}\label{eq_app_limits_hatted_DE}\end{equation}
where
\begin{align}
  \hat{Q} &= 2 \hat{\Sigma}_{+}^{2} + \hat{R}^{2} + \case{1}{6}
    \hat{\mathcal{H}}^{2} + \case{1}{2}(3\gamma-2) \hat{\Omega},
    \label{eq_app_limits_Q_hat} \\
  \hat{\Omega} &= 1 - \hat{\Sigma}_{+}^{2} - \hat{R}^{2} - 
    \case{1}{6} \hat{\mathcal{H}}^{2} .  \label{eq_app_limits_Omega_hat}
\end{align}
One can also form an auxilliary DE for $\hat{\Omega}$ using
\eref{eq_app_limits_hatted_DE} and \eref{eq_app_limits_Omega_hat} to find that
\begin{equation}
  \hat{\Omega} ' = [ 2\hat{Q} - (3\gamma-2)] \hat{\Omega} . 
    \label{eq_app_limits_Omega_DE}
\end{equation}
We consider the state space $S$ of the DE~\eref{eq_app_limits_hatted_DE}
defined by the inequalities
\begin{equation}
  \hat{R} > 0, \qquad \hat{\mathcal{H}} > 0, \qquad \hat{\Omega} > 0.
    \label{eq_app_limits_statespaceS}
\end{equation}
These inequalities in conjunction with \eref{eq_app_limits_Omega_hat} imply
that the state space $S$ is the interior of one quarter of an ellipsoid.
Understanding the dynamics on the two-dimensional invariant sets
$S_{\hat{\Omega}}$, $S_{\hat{R}}$ and $S_{\hat{\mathcal{H}}}$, the closure of
their union defining the boundary of $S$, will be crucial in our analysis.
These sets are defined by the following restrictions:
\begin{align*}
  S_{\hat{\Omega}} &:\quad \hat{\Omega} = 0, \quad \hat{R} > 0, \quad
    \hat{\mathcal{H}} > 0 , \\
  S_{\hat{R}} &:\quad \hat{R} = 0, \quad \hat{\mathcal{H}} > 0, \quad
    \hat{\Omega} > 0 , \\
  S_{\hat{\mathcal{H}}} &:\quad \hat{\mathcal{H}} = 0, \quad \hat{R} > 0,
    \quad \hat{\Omega} > 0 .
\end{align*}

The DE~\eref{eq_app_limits_hatted_DE} admits a positive monotone function
\begin{equation}
  Z = \frac{ \hat{\Omega}^{3} }{ \hat{R}^{4} \hat{\mathcal{H}}^{2} },
    \label{eq_app_limits_Z}
\end{equation}
which satisfies
\begin{equation}
  Z' = 3(4-3\gamma)Z   \label{eq_app_limits_Zprime}
\end{equation}
on the set $S$. Thus, if $\gamma \not= \frac{4}{3}$ there are no equilibrium
points, periodic orbits and homoclinic orbits in $S$ (see \cite{WE},
proposition~4.2). It is immediate upon integrating \eref{eq_app_limits_Zprime}
and using the boundedness of $\hat{R}$ and $\hat{\mathcal{H}}$ that for any
$\hat{\vec{x}} \in S$
\begin{alignat}{3}
  \omega(\hat{\vec{x}}) &\subseteq \bar{S}_{\hat{R}} \cup
    \bar{S}_{\hat{\mathcal{H}}}, &\qquad \text{if} \quad
    \case{2}{3} < \gamma < \case{4}{3} , \label{eq_app_limits_omega1} \\
  \omega(\hat{\vec{x}}) &\subseteq \bar{S}_{\hat{\Omega}}, &\qquad \text{if}
    \quad \case{4}{3} < \gamma \leq 2 . \label{eq_app_limits_omega2}
\end{alignat}

\begin{figure} \small
  \begin{center}
  \includegraphics[scale=0.65]{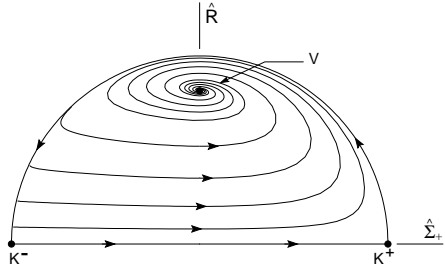}
  \caption{\label{fig_Omega0}Orbits in the invariant set $S_{\hat{\Omega}}$.}
  \end{center}
\end{figure}

We now consider the case $\frac{4}{3} < \gamma \leq 2$. The flow on the
invariant set $S_{\hat{\Omega}}$ is depicted in figure~\ref{fig_Omega0},
which shows the projection of the surface $\hat{\Omega} = 0$ onto the
$\hat{\Sigma}_{+}\hat{R}$-plane. The essential features are the existence of
three equilibrium points
\begin{align*}
    \mathsf{K}^{\pm}: &\quad (\hat{\Sigma}_{+},\hat{R},\hat{\mathcal{H}}) = 
      (\pm 1, 0, 0), \\
    \mathsf{V}: &\quad (\hat{\Sigma}_{+},\hat{R},\hat{\mathcal{H}}) = 
      \left( 0, \sqrt{\case{2}{3}}, \sqrt{2} \right),
\end{align*}
in which $\mathsf{K}^{\pm}$ lie on the boundary of $S_{\hat{\Omega}}$, and
the fact that there are no periodic orbits on $S_{\hat{\Omega}}$. The latter
can be established by the existence of a Dulac function $\lambda$ on
$S_{\hat{\Omega}}$ given by
$$
  \lambda = \hat{R}^{-3} ( 1 - \hat{\Sigma}_{+}^{2} - \hat{R}^{2} )^{-2}
$$
(see \cite{WE}, theorem~4.6, pg.~94). Thus, the only potential $\omega$-limit
sets in $S_{\hat{\Omega}}$ are the equilibrium points $\mathsf{K}^{\pm}$,
$\mathsf{V}$ and the heteroclinic sequence $(\mathsf{K}^{-} \rightarrow
\mathsf{K}^{+} \rightarrow \mathsf{K}^{-})$ and hence for any $\hat{\vec{x}}
\in S_{\hat{\Omega}}$, the $\omega$-limit set is one of these four candidates.
The point $\mathsf{K}^{+}$ can be excluded since it is a local source in $S$.
Moreover, the point $\mathsf{K}^{-}$ can be excluded by considering the
evolution equation for $\hat{\mathcal{H}}$, which is of the form
$$
  \hat{\mathcal{H}}' = h(\hat{\Sigma}_{+},\hat{R},\hat{\mathcal{H}})
    \hat{\mathcal{H}}.
$$
Since $h(\mathsf{K}^{-}) = h(-1,0,0) = 3$ and $\hat{\mathcal{H}} = 0$ at
$\mathsf{K}^{-}$, it follows that $\lim_{\tau \rightarrow +\infty}
\hat{\mathcal{H}} \not= 0$ and hence that an orbit in $S$ cannot be future
asymptotic to $\mathsf{K}^{-}$. This leaves the equilibrium point $\mathsf{V}$
and the heteroclinic sequence $(\mathsf{K}^{-} \rightarrow \mathsf{K}^{+}
\rightarrow \mathsf{K}^{-})$ as the remaining candidates for the
$\omega$-limit set in $S_{\hat{\Omega}}$. The latter can be excluded since
$\mathsf{V}$ is a local sink in $S$ and hence $\omega(\hat{\vec{x}}) =
\mathsf{V}$ for any $\hat{x} \in S_{\hat{\Omega}}$. On account of
\eref{eq_app_limits_omega2}, we thus conclude that for any
$\hat{\vec{x}} \in S$,
\begin{equation}
  \lim_{\tau \rightarrow +\infty} (\hat{\Sigma}_{+}, \hat{R},
    \hat{\mathcal{H}}) = \left( 0, \sqrt{\case{2}{3}}, \sqrt{2} \right) ,
      \qquad \text{if} \quad \case{4}{3} < \gamma \leq 2
      \label{eq_app_limits_caseI}
\end{equation}

The case $\frac{2}{3} < \gamma < \frac{4}{3}$ can be treated in a similar
fashion by analyzing the dynamics on the invariant sets $S_{\hat{R}}$ and
$S_{\hat{\mathcal{H}}}$. It follows that
\begin{equation}
  \lim_{\tau \rightarrow +\infty} (\hat{\Sigma}_{+}, \hat{R},
    \hat{\mathcal{H}}) = (0, 0, 0) , \qquad \text{if} \quad \case{2}{3}
      < \gamma \leq \case{4}{3} .  \label{eq_app_limits_caseII}
\end{equation}

\begin{figure} \small
  \begin{center}
  \includegraphics[scale=0.6]{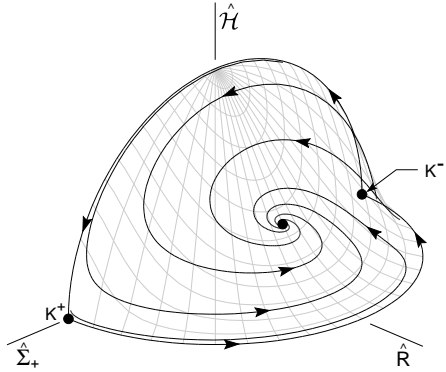}
  \caption{\label{fig_gamma_fourthirds}Orbits in the invariant set
    $\hat{\Omega}^{3}/( \hat{R}^{4} \hat{\mathcal{H}}^{2} ) = k$ with $k > 0$
    and $\gamma = \frac{4}{3}$.}
  \end{center}
\end{figure}

Finally, we consider the case $\gamma = \frac{4}{3}$. We first observe that
the $\hat{\Omega}$ evolution equation~\eref{eq_app_limits_Omega_DE} restricted
to a radiation fluid reduces to
$$
  \hat{\Omega}' = 2 \hat{\Sigma}_{+}^{2} \hat{\Omega}.
$$
It follows immediately from the LaSalle invariance principle (see \cite{WE},
theorem~4.11, pg.~103) that
\begin{equation}
  \omega(\hat{\vec{x}}) \subset \{ \hat{\vec{x}} \:|\: \Sigma_{+} = 0 \},
    \label{eq_app_limits_radC}
\end{equation}
for any $\hat{\vec{x}} \in S$. By~\eref{eq_app_limits_Zprime} the function $Z$
defined in \eref{eq_app_limits_Z} describes a conserved quantity
\begin{equation}
  \frac{ \hat{\Omega}^{3} }{ \hat{R}^{4} \hat{\mathcal{H}}^{2} } = k,
    \label{eq_app_limits_Zconserved}
\end{equation}
where $k > 0$ is a constant that depends on the initial condition. We see
that for all $k > 0$ the surfaces described by \eref{eq_app_limits_Zconserved}
foliate the state space $S$ and intersect the boundary $\hat{\Omega} = 0$ at
$\hat{R} = 0$ and $\hat{\mathcal{H}} = 0$
(see figure~\ref{fig_gamma_fourthirds}). When $\gamma = \frac{4}{3}$, the
DE~\eref{eq_app_limits_hatted_DE} has a line $\mathsf{L}$ of equilibrium
points given by
$$
  \mathsf{L}: \quad (\hat{\Sigma}_{+}, \hat{R}, \hat{\mathcal{H}}) =
    \left(0, \sqrt{\case{2}{3}}\, k, \sqrt{2}\, k \right), \quad k \in (0,1).
$$
It can show that for each $k > 0$, the two-dimensional invariant set defined
by \eref{eq_app_limits_Zconserved} intersects the line $\mathsf{L}$ at
precisely one point. Since this unique point of intersection is the only
equilibrium point on this invariant set which satisfies
$\hat{\Sigma}_{+} = 0$, it follows from the restriction
\eref{eq_app_limits_radC} that any solution in $S$ satisfies
\begin{equation}
  \lim_{\tau \rightarrow +\infty} (\hat{\Sigma}_{+}, \hat{R},
    \hat{\mathcal{H}}) =  \left(0, \sqrt{\case{2}{3}}\, k, \sqrt{2}\, k
    \right) , \qquad \text{if} \quad \gamma = \case{4}{3} ,
    \label{eq_app_limits_caseIII}
\end{equation}
where $k \in (0,1)$ is a constant which depends on the initial condition.


\stepcounter{subsection}
\subsection*{Appendix~\thesubsection}

We now apply theorem~\ref{theorem_SY} using the results of appendix~B.1
to prove that
$$
  \lim_{\tau \rightarrow +\infty} \bar{\vec{x}} = \vec{a},
$$
where $\bar{\vec{x}} = (\bar{\Sigma}_{+},\bar{R},\bar{\mathcal{H}})$ and
$\vec{a}$ is given by the right-hand sides of \eref{eq_app_limits_caseI},
\eref{eq_app_limits_caseII} and \eref{eq_app_limits_caseIII}, considering
the three cases $\frac{2}{3} < \gamma < \frac{4}{3}$, $\gamma = \frac{4}{3}$
and $\frac{4}{3} < \gamma \leq 2$ simultaneously.

We begin by defining the subset $D$ in theorem~\ref{theorem_SY} by
$$
  \Sigma_{+}^{2} + R^{2} + \case{1}{6} \mathcal{H}^{2} < 1.
$$
We now verify the hypotheses $H_{1}$ and $H_{2}$. Firstly, let
$\vec{w} : [\tau_{0},+\infty) \rightarrow D$ be any $C^{0}[\tau_{0},\infty)$
function. Since $\lim_{\tau \rightarrow +\infty} M(\tau) = 0$ it follows
immediately from \eref{eq_limits_g} that
$$
  \lim_{\tau \rightarrow +\infty} \vec{g}(\vec{w}(\tau),\tau) =
  \lim_{\tau \rightarrow +\infty} M(\tau) \left( B_{\bar{\Sigma}_{+}},
    \bar{R} B_{\bar{R}}, \bar{\mathcal{H}} B_{\bar{\mathcal{H}}} \right)
    \Big|_{\bar{\vec{x}} = \vec{w}(\tau)} = \mathbf{0},
$$
showing that $H_{1}$ is satisfied. Secondly, $H_{2}$ is satisfied
since the variables $\bar{\Sigma}_{+}$, $\bar{R}$ and $\bar{\mathcal{H}}$ are
bounded for all $\tau \geq \tau_{0}$ with $\tau_{0}$ sufficiently large.
Therefore, since
$$
  \lim_{\tau \rightarrow +\infty} \hat{\vec{x}}(\tau) = \vec{a}
$$
for all initial conditions $\hat{\vec{x}}(\tau_{0})$ in $D$
(see~\eref{eq_app_limits_caseI}, \eref{eq_app_limits_caseII} or
\eref{eq_app_limits_caseIII}), theorem~\ref{theorem_SY} implies that
\begin{equation}
  \lim_{\tau \rightarrow +\infty} \bar{\vec{x}}(\tau) = \vec{a}
    \label{eq_app_limits_SYmainresult}
\end{equation}
for all initial conditions $\bar{\vec{x}}(\tau_{0})$ in $D$.

Finally, we need to show that any initial condition $\vec{x}(\tau_{0}) =
(\Sigma_{+},R,\mathcal{H}) \big|_{\tau=\tau_{0}}$, $M(\tau_{0})$,
$\psi(\tau_{0})$ for the DE~\eref{eq_eveqn_full_DE}, subject to $\Omega > 0$
and \eref{eq_eveqn_BMVII0b}, determines an initial condition
$\bar{\vec{x}}(\tau_{0})$ in $D$ for the DE~\eref{eq_limits_nonautonomous_DE},
so that \eref{eq_app_limits_SYmainresult} is satisfied. Indeed, since
$\lim_{\tau \rightarrow +\infty} \psi = +\infty$, we can without loss of
generality restrict the initial condition $\psi(\tau_{0})$ to be a multiple of
$\pi$. This requirement can be achieved by simply following the solution
determined by the original initial condition until this condition is
satisfied. It follows from this condition, in conjunction with
\eref{eq_limits_cov} and the restriction $\Omega > 0$ applied to
\eref{eq_eveqn_Omega}, that
$$
  \left. \left( \bar{\Sigma}_{+}^{2} + \bar{R}^{2} + \case{1}{6}
    \bar{\mathcal{H}}^{2} \right) \right|_{\tau = \tau_{0}} = \left.
    \left( \Sigma_{+}^{2} + R^{2} + \case{1}{6} \mathcal{H}^{2} \right)
    \right|_{\tau=\tau_{0}} < 1,
$$
so that $\bar{\vec{x}}(\tau_{0}) \in D$.


\stepcounter{subsection}
\subsection*{Appendix~\thesubsection}

We now provide the proof of \eref{eq_limits_MoverR} for the case $\frac{2}{3}
< \gamma < \frac{4}{3}$, which gives the limit of the ratio $R/M$ at late
times. In analogy to \eref{eq_limits_cov}, we define a variable $\bar{M}$ by
\begin{equation}
  \bar{M} = M \left( 1 + \case{1}{4} R^{2} M \sin 2 \psi \right).
    \label{eq_RoverM_Mbar}
\end{equation}
It follows from \eref{eq_eveqn_full_DE} that the evolution equation for
$\bar{M}$ is of the form
\begin{equation}
  \bar{M}' = -(\bar{Q} + 2 \bar{\Sigma}_{+} + M B_{\bar{M}}) \bar{M},
    \label{eq_RoverM_eveqn_Mbar}
\end{equation}
where $B_{\bar{M}}$ is a bounded function for $\tau$ sufficiently large.
By using \eref{eq_limits_barred_DE} we obtain
\begin{equation}
  \left( \frac{\bar{R}}{\bar{M}} \right)' = \left[ 3(\gamma-1) + 
    h(\bar{\vec{x}},M,\psi) \right] \left( \frac{\bar{R}}{\bar{M}} \right),
    \label{eq_RoverM_eveqn_RoverM}
\end{equation}
where
$$
  h(\bar{\vec{x}},M,\psi) = 2 - 3 \gamma + 2 \bar{Q} + 3 \bar{\Sigma}_{+} + 
    M B_{*}
$$
and $B_{\ast}$ is a bounded function for $\tau$ sufficiently large. It follows
from \eref{eq_limits_M}, \eref{eq_limits_cov}, \eref{eq_limits_Q} and
theorem~\ref{theorem_limits} that $\lim_{\tau \rightarrow +\infty}
h(\bar{\vec{x}},M,\psi) = 0$. Consequently, \eref{eq_RoverM_eveqn_RoverM}
implies that
\begin{align*}
  \frac{\bar{R}}{\bar{M}} &= \bigO\left( \rme^{[3(\gamma-1)+\delta]\tau}
    \right) , \qquad \text{if} \quad \case{2}{3} \leq \gamma < 1 , \\
  \frac{\bar{M}}{\bar{R}} &= \bigO\left( \rme^{[3(1-\gamma)+\delta]\tau}
    \right) , \qquad \text{if} \quad 1 < \gamma < \case{4}{3} ,
\end{align*}
as $\tau \rightarrow +\infty$ for any $\delta > 0$. Therefore, on account of
\eref{eq_RoverM_Mbar} and \eref{eq_limits_cov},
$$
  \lim_{\tau \rightarrow +\infty} \frac{R}{M} = \begin{cases}
    0, &\text{if}\quad \frac{2}{3} \leq \gamma < 1,\\
    +\infty, &\text{if}\quad 1 < \gamma < \frac{4}{3}. \end{cases}
$$

It remains to deduce the limit of $R/M$ as $\tau \rightarrow +\infty$ for
the case $\gamma = 1$. To proceed we compute the asymptotic form of
$R$ and $M$ as $\tau \rightarrow +\infty$. The calculation parallels that for
the non-magnetic Bianchi VII$_{0}$ models detailed in appendix~B of
\cite{WHU99}. It follows that any solution of the DE~\eref{eq_eveqn_full_DE}
subject to the restrictions \eref{eq_eveqn_BMVII0b} with
$\frac{2}{3} < \gamma < \frac{4}{3}$ satisfies
\begin{align*}
  \Sigma_{+} &= \frac{2(C_{\mathcal{H}}^{2} - 3 C_{R}^{2})}{3(3\gamma-2)} \, 
    \rme^{(3\gamma-4)\tau} \left[ 1 + \bigO(\rme^{-b\tau}) \right], \\
  R &= C_{R}\, \rme^{1/2\,(3\gamma-4)\tau} \left[ 1 + 
    \bigO(\rme^{-b\tau}) \right], \\
  \mathcal{H} &= C_{\mathcal{H}}\, \rme^{1/2\,(3\gamma-4)\tau} 
    \left[ 1 + \bigO(\rme^{-b\tau}) \right], \\
  M &= C_{M}\, \rme^{1/2\,(2-3\gamma)\tau}
    \left[ 1 + \bigO(\rme^{-b\tau}) \right],
\end{align*}
as $\tau \rightarrow +\infty$, for some constant $b > 0$, where $C_{R}$,
$C_{\mathcal{H}}$ and $C_{M}$ are positive constants which depend on the
initial conditions. Therefore,
$$
  \frac{R}{M} = \frac{C_{R}}{C_{M}}\, \rme^{3(\gamma-1)\tau}
    \left[ 1 + \bigO(\rme^{-b\tau}) \right]
$$
as $\tau \rightarrow +\infty$ and hence
$$
  \lim_{\tau \rightarrow +\infty} \frac{R}{M} = \frac{C_{R}}{C_{M}} \not= 0 ,
    \qquad \text{if} \quad \gamma = 1.
$$


\stepcounter{section}
\section*{Appendix~\thesection}

In this appendix we give an expression for the Weyl curvature parameter
$\mathcal{W}$ in terms of the Hubble-normalized variables $\Sigma_{+}$,
$R$, $\mathcal{H}$, $M$ and $\psi$. Let $E_{\alpha\beta}$ and
$H_{\alpha\beta}$ be the components of the electric and magnetic parts of the
Weyl tensor relative to the group invariant frame with $\vec{e}_{0} = \vec{u}$.
It follows that $E_{\alpha\beta}$ and $H_{\alpha\beta}$ are diagonal and
trace-free and hence they each have two independent components. In analogy
with \eref{eq_eveqn_Npm} we define
\begin{equation}\begin{split}
  \mathcal{E}_{+} = \case{1}{2}(\mathcal{E}_{22}+\mathcal{E}_{33}), &\qquad
  \mathcal{E}_{-} = \case{1}{2\sqrt{3}}(\mathcal{E}_{22}-\mathcal{E}_{33}), \\
  \mathcal{H}_{+} = \case{1}{2}(\mathcal{H}_{22}+\mathcal{H}_{33}), &\qquad
  \mathcal{H}_{-} = \case{1}{2\sqrt{3}}(\mathcal{H}_{22}-\mathcal{H}_{33}),
\end{split}\label{eq_app_Weyl_a}\end{equation}
where $\mathcal{E}_{\alpha\beta}$ and $\mathcal{H}_{\alpha\beta}$ are the
dimensionless counterparts of $E_{\alpha\beta}$ and $H_{\alpha\beta}$,
defined by
\begin{equation}
  \mathcal{E}_{\alpha\beta} = \frac{E_{\alpha\beta}}{H^{2}}, \qquad
  \mathcal{H}_{\alpha\beta} = \frac{H_{\alpha\beta}}{H^{2}}.
    \label{eq_app_Weyl_b}
\end{equation}
It follows from \eref{eq_limits_Weyl_defn}, \eref{eq_app_Weyl_a} and
\eref{eq_app_Weyl_b} that
\begin{equation}
  \mathcal{W}^{2} = \mathcal{E}_{+}^{2} + \mathcal{E}_{-}^{2}
    + \mathcal{H}_{+}^{2} + \mathcal{H}_{-}^{2}. \label{eq_app_Weyl_c}
\end{equation}
Equations (1.101) and (1.102) in \cite{WE} for $E_{\alpha\beta}$ and
$H_{\alpha\beta}$, in conjunction with the frame choice detailed in section~2
of \cite{LKW95} and equations \eref{eq_app_Weyl_a} and \eref{eq_eveqn_cov} in
the present paper lead to
\begin{equation}\begin{split}
  \mathcal{E}_{+} &= \Sigma_{+}(1+\Sigma_{+}) + \case{1}{2}
    R^{2}(1-3 \cos 2 \psi) - \case{1}{6} \mathcal{H}^{2} , \\
  \mathcal{H}_{+} &= -\case{3}{2} R^{2} \sin 2 \psi , \\
  \mathcal{E}_{-} &= \frac{2R}{M} \left[ \sin \psi + \case{1}{2}M( 1 -
    2 \Sigma_{+}) \cos \psi \right] , \\
  \mathcal{H}_{-} &= \frac{2R}{M} \left[ -\cos \psi - \case{3}{2} M \Sigma_{+}
    \sin \psi \right] .
\end{split}\label{eq_app_Weyl_explicit}\end{equation}


\end{document}